\documentclass[conference,10pt,a4paper]{IEEEtran}

\usepackage[noadjust]{cite}
\usepackage{graphicx}
\usepackage{amsmath}
\usepackage{amsfonts}
\usepackage{algorithm}
\usepackage{algpseudocode}
\usepackage{amsthm}
\usepackage{bm}
\usepackage{xcolor}

\usepackage{tikz,pgfplots}

\usepackage[caption=false, font=footnotesize]{subfig}

\theoremstyle{plain}

\theoremstyle{definition}

\theoremstyle{remark} 
\newtheorem{remark}{Remark}

\IEEEoverridecommandlockouts

\begin{document}



\title{Decoding Reed--Muller Codes Using \\ Minimum-Weight Parity Checks}


\author{\IEEEauthorblockN{Elia Santi\IEEEauthorrefmark{1},  Christian
H\"{a}ger\IEEEauthorrefmark{2}\IEEEauthorrefmark{3}, and Henry D.
Pfister\IEEEauthorrefmark{3}}

\thanks{This work was done while E.~Santi was visiting Duke
University, Durham, North Carolina. The work of C.~H\"{a}ger was
supported by the European Union's Horizon 2020 research and innovation
programme under the Marie Sk\l{}odowska-Curie grant No.~749798. The
work of H.~D.~Pfister was supported in part by the NSF under Grant
No.~1718494.  Any opinions, findings, conclusions, and recommendations
expressed in this material are those of the authors and do not
necessarily reflect the views of the sponsors. Corresponding author
e-mail: \texttt{henry.pfister@duke.edu}.}

\IEEEauthorblockA{\IEEEauthorrefmark{1}Department of Information
Engineering, University of Parma, Parma, Italy}
\IEEEauthorblockA{\IEEEauthorrefmark{2}Department of Electrical
Engineering, Chalmers University of Technology, Gothenburg, Sweden}
\IEEEauthorblockA{\IEEEauthorrefmark{3}Department of Electrical and
Computer Engineering, Duke University, Durham, North Carolina}
}


\maketitle

\newcommand{\markred}[1]{{\color{red} #1}}
\newcommand{\markblue}[1]{{\color{blue} #1}}

\begin{abstract}
	Reed--Muller (RM) codes exhibit good performance under
	maximum-likelihood (ML) decoding due to their highly-symmetric
	structure. In this paper, we explore the question of whether the
	code symmetry of RM codes can also be exploited to achieve near-ML
	performance in practice. The main idea is to apply iterative
	decoding to a highly-redundant parity-check (PC) matrix that
	contains only the minimum-weight dual codewords as rows.  As
	examples, we consider the peeling decoder for the binary erasure
	channel, linear-programming and belief propagation (BP) decoding for
	the binary-input additive white Gaussian noise channel, and
	bit-flipping and BP decoding for the binary symmetric channel. For
	short block lengths, it is shown that near-ML performance can indeed
	be achieved in many cases. We also propose a method to tailor the PC
	matrix to the received observation by selecting only a small
	fraction of useful minimum-weight PCs before decoding begins. This
	allows one to both improve performance and significantly reduce
	complexity compared to using the full set of minimum-weight PCs. 
\end{abstract}






\IEEEpeerreviewmaketitle

\section{Introduction}


Recently, the 5G cellular standardization process focused on
error-correcting codes and decoders that are nearly optimal for short
block lengths (e.g., rate-$1/2$ binary codes with lengths from $128$
to $512$). Promising contenders include modified polar codes,
low-density parity-check (LDPC) codes, and tail-biting convolutional
codes~\cite{Tal-it15,Liva-arxiv16,vanWonterghem-arxiv16}. These
results also show that short algebraic codes such as Reed--Muller (RM)
and extended BCH (eBCH) codes under ML decoding tie or outperform all
the other choices. ML performance can be approached with methods based
on ordered-statistics decoding~\cite{Fossorier-it95} such as most-reliable basis (MRB) decoding \cite{Dorsch-it74}. Depending on the code and
decoder parameters, the complexity of these methods range from
relatively practical to extremely complex. 

Motivated by the good performance of RM codes under ML decoding and
the proof that RM codes achieve capacity on the binary erasure channel
(BEC) \cite{Kudekar-it17}, we revisit the question of whether code
symmetry can be used to achieve near-ML performance in practice.  This
question, in various forms, has been addressed by several groups over
the past years~\cite{Halford-isit06,Hehn-isit07,Hehn-com10}.  In
general, one applies a variant of belief-propagation (BP) decoding to
the Tanner graph defined by a redundant parity-check (PC) matrix for
the code. Often, the redundancy in the PC matrix is derived from the
symmetry (i.e., automorphism group) of the code. Methods based on
redundant PC matrices are also related to earlier approaches that
adapt the PC matrix during
decoding~\cite{Jiang-it06,Kothiyal-comlett05,Taghavi-it08}. 

Highly-symmetric codes such as RM codes can have a very large number
of minimum-weight (MW) PCs. The main contribution of this paper is to
show that this large set of MWPCs can be exploited to provide near-ML
performance with several well-known iterative decoding schemes. In
particular, we consider the peeling decoder (PD) for the BEC,
linear-programming (LP) and BP decoding for the binary-input additive
white Gaussian noise (AWGN) channel, and bit-flipping (BF) and BP
decoding for the binary symmetric channel (BSC). It is worth noting
that the idea of using all MWPCs for decoding appears
in~\cite{Bossert-it86,Lucas-jsac98}, even before the rediscovery of iterative
decoding. Their decoding algorithms are variations of the BF
algorithm~\cite{RU-2008} applied to a redundant PC matrix. As we will
see, the BF algorithms they propose work remarkably well for the BSC,
but result in a performance loss compared to LP and BP decoding for
the AWGN channel.

Using all available MWPCs for decoding can become quite complex, e.g.,
the rate-1/2 RM code of length 128 has 94,488 MWPCs. To address this
problem, we propose a method to select only a small fraction of useful
MWPC based on the channel reliabilities. We exploit the fact that,
given any small set of bit positions, there exist efficient methods to
find a MWPC that contains the chosen bits. This process is iterated to
design a redundant PC matrix that is tailored to the received
observation from the channel. The resulting PC matrix can allow one to
both improve performance (by reducing cycles in the Tanner graph) and
reduce complexity. We stress that this approach works by adapting the
PC matrix \emph{before} decoding begins. Thus, it is closer in spirit
to MRB decoding than to the adaptive methods employed
by~\cite{Jiang-it06,Kothiyal-comlett05,Taghavi-it08}. 

\section{Background on Reed--Muller Codes}
\label{sec:bg}
\newcommand{\dmin}{\ensuremath{d_\text{min}}}


RM codes were introduced by Muller in~\cite{Muller-ire54}. We use
$\mathrm{RM}(r,m)$ to denote the $r$-th order RM code of length
$n=2^m$, where $0 \leq r \leq m$. Each codeword in $\mathrm{RM}(r,m)$
is defined by evaluating a multivariate polynomial $f\in \mathbb{F}_2
[x_1,\ldots,x_m]$ of degree at most $r$ at all points in
$\mathbb{F}_2^m$~\cite{MacWilliams-1977}. The code $\mathrm{RM}(r,m)$
has minimum distance $\dmin = 2^{m-r}$ and dimension $k = \binom{m}{0}
+ \cdots + \binom{m}{r}$. 

\subsection{Number of Minimum-Weight Parity Checks}

For a binary linear code, the codewords of the dual code define all
valid rows of the PC matrix. Since the dual code of $\mathrm{RM}(r,m)$
is $\mathrm{RM}(m-r-1,m)$, the MWPCs of $\mathrm{RM}(r,m)$ thus have
weight $2^{m - (m-r-1)} = 2^{r+1}$. In order to determine the number
of MWPCs for $\mathrm{RM}(r,m)$, one may use the fact that each MW
codeword of $\mathrm{RM}(m-r-1,m)$ is the indicator vector of an
$(r+1)$-dimensional affine subspace of $\mathbb{F}_2^m$
\cite{MacWilliams-1977}. Based on this, one can show that the number
of MWPCs is given by \cite{MacWilliams-1977}
\begin{align}
 F(r,m) = 2^{m-r-1} \prod_{i=0}^r \frac{2^{m-i} - 1}{2^{r+1-i} -1}.
\end{align}
For example, the
$[128,64,16]$ code\footnote{A linear code is called an
$[n, k, d]$ code if it has length $n$, dimension $k$, and
minimum-distance $d$.} $\mathrm{RM}(3,7)$ has 94,488 weight-16 PCs.






\subsection{Generating Minimum-Weight Parity Checks}

The connection between MWPCs and affine subspaces also provides an
efficient method for generating a MWPC that is connected to any given
set of $r+2$ codeword bits. In particular, one can simply complete the
affine subspace containing the chosen $r+2$ points. If the chosen set
of points is not affinely independent, then one can extend the set to
define an $(r+1)$-dimensional affine subspace. This procedure is
described in Algorithm~\ref{alg:rm2chk}. The algorithm will be used to
construct a PC matrix for $\mathrm{RM}(r,m)$ that is tailored to a
particular received sequence. This procedure is described in the next
section. 


\section{Parity-Check Matrix Adaptation}
\label{sec:adaptation}

\newcommand{\Hfull}{\boldsymbol{H}_{\text{full}}}
\newcommand{\Hsub}{\boldsymbol{H}_{\text{sub}}}


The PC matrix containing all MW dual codewords as rows is denoted by
$\Hfull$. This matrix can be used directly for iterative decoding,
e.g., BP decoding.  In general, however, the decoding complexity for the
considered iterative schemes scales linearly with the number of rows
in the PC matrix.  Thus, depending on the RM code, decoding based on
the full matrix $\Hfull$ may result in high complexity. 

On the other hand, not all the rows of $\Hfull$ are equally useful in
the decoding of a particular received observation vector
$\boldsymbol{y} = (y_1, \dots, y_n)^\top$. For instance, if all the
bits involved in a given PC are relatively unaffected by the channel,
the associated PC would be uninformative for the decoding
process. Therefore, our approach to reduce complexity is to pick only
the rows of $\Hfull$ that are expected to be useful for the decoding.
The choice of rows is based on $\boldsymbol{y}$ and the resulting PC
matrix containing the subset of rows is denoted by $\Hsub$.  

\begin{algorithm}[t]

\caption{\label{alg:rm2chk} For $\mathrm{RM}(r,m)$, generate MWPC $\boldsymbol{w} \in
\mathbb{F}_2^{n}$  with ones in bit positions $i_1,\dots,i_{r+2} \in \{1,\dots,n\}$, where $n=2^m$ }

\begin{algorithmic}[1]
\Statex
\State Let row-vector $\bm{v}_j \in \mathbb{F}_2^m$ be the binary expansion of $i_j -1$
\State Form matrix $\bm{A}\in \mathbb{F}_2^{(r+1)\times m}$ with rows $\bm{a}_j = \bm{v}_j \oplus \bm{v}_{r+2}$
\State $\bm{B} \; \leftarrow$ reduced row echelon form of $\bm{A}$
\While{$\bm{B}$ contains all-zero rows}
   \State In first column that is not equal to a unit vector, add a one at row position of the first all-zero row
\EndWhile
\State Initialize $\bm{w} \in \mathbb{F}_2^n$ to the all-zero vector
\ForAll { $l\in \{0,\dots,2^{r+1}-1\}$}
\State  $\bm{u}_l \; \leftarrow $ $m$-bit binary expansion of $l$
 \State $\bm{z}_l \leftarrow \bm{u}_l \bm{B} \oplus \bm{v}_{r+2}$
 \State $s_l \; \leftarrow $ integer represented by binary expansion $\bm{z}_l$
 \State $\bm{w}_{s_l +1} \leftarrow 1$
\EndFor
\end{algorithmic}
\end{algorithm}

%

\subsection{General Idea}

In order to illustrate the general idea, suppose the codeword bits are
transmitted through a channel and the received values are classified
either as \emph{good} or \emph{bad}. For example, on the BEC an
unerased bit would be labeled good while an erased bit would be called
bad. Then, Algorithm~\ref{alg:rm2chk} can be used to generate a MWPC
that contains one bad bit of interest, $r+1$ randomly chosen good
bits, and some set of $2^{r+1}-r-2$ other bits. From an
information-theoretic point of view, this MWPC is expected to provide
more information about the bad bit of interest than a random MWPC
because it involves a guaranteed number of $r+1$ good bits. Repeating
this process allows one to generate a PC matrix for $\mathrm{RM}(r,m)$
that is biased towards informative MWPCs.

\subsection{Reliability Criterion}

As a first step, the bit positions $I = \{ 1,2,3,\dots,n \}$ are
divided into two disjoint sets $G$ and $B$ based on their reliability.
To that end, one first computes the vector of log-likelihood ratios
(LLRs) $\boldsymbol{\gamma} = (\gamma_1, \dots, \gamma_n)^\top \in
\mathbb{R}^n$ based on the received vector $\boldsymbol{y}$. The
vector $\boldsymbol{\gamma}$ is then sorted, i.e., $(t_1,\dots,t_{n})$
is a permutation of bit indexes such that $i>j \Rightarrow
|\gamma_{t_i}| \geq |\gamma_{t_j}|$. We then set $G=\{ t_k\in I : k
\leq f n \}$ and $B=I-G$, where $0 \leq f \leq 1$ is a tunable
parameter and $f n$ is assumed
to be an integer. 

\begin{remark}
	The LLR sorting can be applied for an arbitrary binary-input
	memoryless channel with the exception of the BSC. The BSC is
	discussed separately in Sec.~\ref{sec:bsc} below.
\end{remark}

\begin{remark}
	The use of sorting may be avoided by instead thresholding the LLRs.
	However, our numerical studies showed that this results in some loss for the short block-lengths we considered. 
\end{remark}

\subsection{Tailoring $\Hsub$ to the Received Vector}
\label{sec:tailor}

The sets of reliable and unreliable bit positions $G$ and $B$ are then
used to generate an overcomplete PC matrix that is tailored to the
received vector $\boldsymbol{y}$. The proposed method is illustrated
in Algorithm 2 below, where $s \in \mathbb{N}$ is the targeted number
of rows of $\Hsub$. Essentially, one iterates through the set of
unreliable bit positions $B$, pairing at each step one unreliable bit
with $r+1$ reliable ones. Based on the resulting set of $r+2$ bit
positions, Algorithm 1 is then used to generate a MWPC (line 5). The
generated MWPC is accepted if it does not already exist in $\Hsub$. We
remark that the if-condition in line 6 of Algorithm 2 can be
implemented very efficiently by applying a hashing function to the
vector $\boldsymbol{w}$ and storing the result in a hashtable. 


%

\begin{algorithm}[t] 
\caption{\label{alg:goodbad} For index sets $G/B$ of good/bad bits, generate a tailored PC matrix $\Hsub$ with $s$ rows}
\begin{algorithmic}[1]
\Statex
\State Initialize $\Hsub$ to an empty matrix 

\While {$\Hsub$ has less than $s$ rows}
\ForAll {$b\in B$}
\State Draw $\{g_1, \dots, g_{r+1}\}$ random positions from $G$ 
\State Generate MWPC $\boldsymbol{w}$ based on $\{b, g_1, \dots, g_{r+1}\}$ 
\If {$\boldsymbol{w}$ is not already a row in $\Hsub$}
\State Append row $\boldsymbol{w}$ to $\Hsub$
\EndIf
\EndFor
\EndWhile
\end{algorithmic}
\end{algorithm}


\section{Decoding Algorithms}

In this section, we briefly review the decoding algorithms that are
used in this paper. 

\subsection{Peeling Decoder}

The PD is an iterative decoder for binary linear codes transmitted over the BEC~\cite{RU-2008}.
It operates on the PC matrix of the code and tracks whether the value of each bit is currently known or unknown.
If there is a PC equation with exactly one unknown bit, then the value of that bit can be computed from the equation and the process is repeated.
Once there is no such PC, the algorithm terminates.

\subsection{Belief Propagation}


BP decoding is an iterative method for decoding binary linear codes transmitted over memoryless channels~\cite{RU-2008}.
It works by passing messages along the edges of the Tanner graph. 
If the graph is a tree, then BP decoding produces optimal decisions.
In general, it is suboptimal and its loss in performance is often attributed to cycles in the graph.

For a code whose Tanner graph has many cycles, it is known that introducing a scaling parameter can improve performance~\cite{Halford-isit06}. 
When using a redundant PC matrix, this can also be motivated by the existence of correlations between input messages to a bit node.
Since BP is based on an independence assumption, these correlations typically cause it to generate bit estimates that are overconfident.
If these messages are represented by LLRs, then this overconfidence can be reduced by scaling messages by a constant less than one.
This approach was also proposed to mitigate the overconfidence associated with min-sum decoding~\cite{Chen2002}.
In this work, the input messages to the bit nodes are scaled by the factor~$w$.

%
%
%
%
%
%

\subsection{Linear-Programming Decoding}
\label{sec:lp}


LP decoding was introduced in \cite{Feldman-it05}.  It is based on
relaxing the ML decoding problem into the linear program $$ \min
\sum_{i=1}^n x_i \gamma_i \\ \;\text{ subject to }\; \boldsymbol{x}\in
\bigcap_{j\in \mathcal{J}} \mathcal{P}_j , $$ where $\mathcal{P}_{j}$
denotes the convex hull of all $\{0,1\}^n$ vectors that satisfy the
$j$-th PC equation.  If the solution vector lies in $\{0,1\}^n$, then
it is the ML codeword.
In theory, a nice property of LP decoding is that the answer is static
and does not depend on the presence of cycles in the Tanner graph.
But, in practice, solving the LP with conventional solvers can be slow
and cycles may affect the convergence speed.

We employ LP decoding using the alternating direction method of multipliers
(ADMM), as proposed in \cite{Barman2013}. The method is based on an
augmented Lagrangian which is parameterized by a tunable scaling
parameter $\mu>0$ \cite[Eq.~(3.2)]{Barman2013}. Then, LP decoding can
be implemented as a message-passing algorithm with an update schedule
similar to BP, where the update rules can be found in
\cite[Sec.~3.1]{Barman2013}. The algorithm stops after $T_\text{max}$
iterations or when a valid codeword is found. 



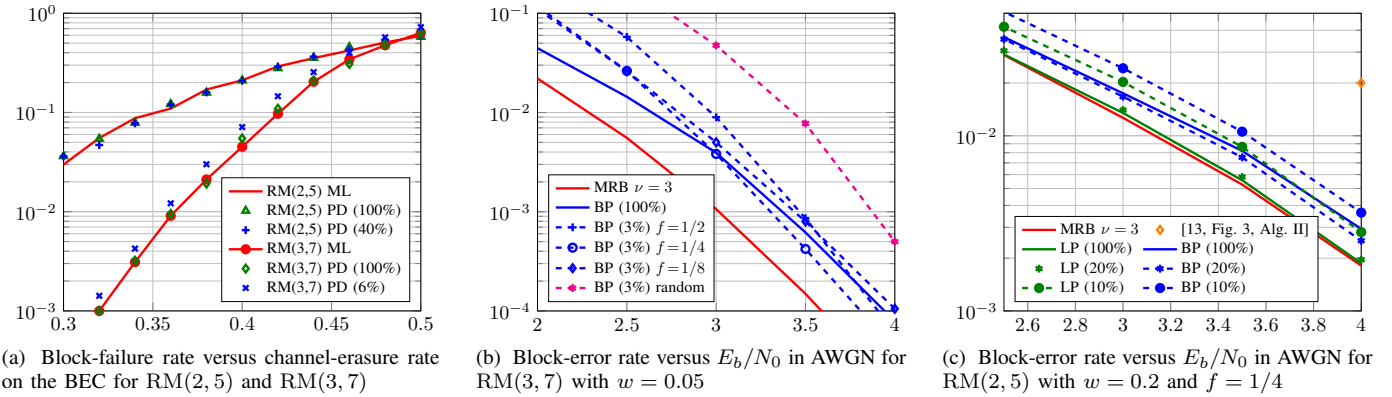
\begin{figure*}
	\centering
	\subfloat[\label{fig:bec12} Block-failure rate versus
	channel-erasure rate on the BEC for $\mathrm{RM}(2,5)$ and $\mathrm{RM}(3,7)$]{\scalebox{0.74}{
%
%
\definecolor{mycolor1}{rgb}{1.00000,0.00000,1.00000}%

\pgfplotsset{
  every axis plot/.append style={very thick}
}

\begin{tikzpicture}

\begin{axis}[%
width=2.5in,
height=2.1in,
at={(1.011in,0.747in)},
scale only axis,
xmin=0.3,
xmax=0.5,
ymode=log,
ymin=0.001,
ymax=1,
yminorticks=true,
axis background/.style={fill=white},
xmajorgrids,
ymajorgrids,
yminorgrids,
legend style={nodes={scale=0.8, transform shape}, at={(0.97,0.03)}, anchor=south east, legend cell align=left, align=left}
]
\addplot [color=red,solid]
  table[row sep=crcr]{%
0.3	0.029875\\
0.32	0.05525\\
0.34	0.0875\\
0.36	0.1095\\
0.38	0.17025\\
0.4	0.211\\
0.42	0.291\\
0.44	0.353\\
0.46	0.4215\\
0.48	0.5075\\
0.5	0.5915\\
};
\addlegendentry{RM(2,5) ML}

\addplot [color=green!50!black, only marks, mark=triangle]
  table[row sep=crcr]{%
0.3	0.03625\\
0.32	0.05425\\
0.34	0.07875\\
0.36	0.1215\\
0.38	0.15825\\
0.4	0.21\\
0.42	0.2795\\
0.44	0.3555\\
0.46	0.4555\\
0.48	0.5185\\
0.5	0.579\\
};
\addlegendentry{RM(2,5) PD (100\%)}

\addplot [color=blue, only marks, mark=+]
  table[row sep=crcr]{%
0.3	0.036\\
0.32	0.047\\
0.34	0.07775\\
0.36	0.12025\\
0.38	0.15925\\
0.4	0.209\\
0.42	0.29\\
0.44	0.3655\\
0.46	0.427\\
0.48	0.539\\
0.5	0.597\\
};
\addlegendentry{RM(2,5) PD (40\%)}

\addplot [color=red, solid, mark=*]
  table[row sep=crcr]{%
0.32	0.001\\ 
0.34	0.0031\\ 
0.36	0.0091\\
0.38	0.0211\\
0.4	    0.0451\\
0.42	0.097\\
0.44	0.204\\
0.46	0.343\\
0.48	0.475\\
0.5	0.6355\\
};
\addlegendentry{RM(3,7) ML}

\addplot [color=green!50!black, only marks, mark=diamond]
  table[row sep=crcr]{%
0.32	0.001\\
0.34	0.0032\\ 
0.36	0.0095\\ 
0.38	0.0192\\ 
0.4	    0.05475\\
0.42	0.109\\
0.44	0.21\\
0.46	0.3085\\
0.48	0.4835\\
0.5	    0.637\\
};
\addlegendentry{RM(3,7) PD (100\%)}

\addplot [color=blue, only marks, mark=x]
  table[row sep=crcr]{%
0.3	0.00047265625\\
0.32	0.00141929791390937\\
0.34	0.00425118629941794\\
0.36	0.0121425199798167\\
0.38	0.0300514168760504\\
0.4	    0.0712223569490155\\
0.42	0.14572192513369\\
0.44	0.255272344698962\\
0.46	0.40016831474858\\
0.48	0.571879674256274\\
0.5	    0.723516282264065\\
};
\addlegendentry{RM(3,7) PD (6\%)}


\end{axis}
\end{tikzpicture}%
}}
	$\quad$
	\subfloat[\label{fig:awgn_opt_f} Block-error rate versus $E_b/N_0$
	in AWGN
	for $\mathrm{RM}(3,7)$ with $w=0.05$]{\scalebox{0.74}{
%
%

\pgfplotsset{
  every axis plot/.append style={very thick}
}

\begin{tikzpicture}

\begin{axis}[%
width=2.5in,
height=2.1in,
scale only axis,
xmin=2,
xmax=4,
xlabel style={font=\small},
ymode=log,
ymin=0.0001,
ymax=0.1,
yminorticks=true,
ylabel style={font=\small},
axis background/.style={fill=white},
title style={font=\bfseries},
xmajorgrids,
ymajorgrids,
yminorgrids,
legend style={nodes={scale=0.8, transform shape}, anchor=south west,at={(0.03,0.03)},legend cell align=left, align=left,/tikz/column 2/.style={column sep=3pt}}
]
\addplot [color=red]
  table[row sep=crcr]{%
2	0.0220348039728752\\
2.5	0.00555361950209024\\
3	0.00106235273475712\\
3.5	0.00014895\\
4	1.625e-05\\
4.5	1.12128146453089e-06\\
};
\addlegendentry{MRB $\nu = 3$}

\addplot [color=blue]
  table[row sep=crcr]{%
2	0.04453125\\
2.5	0.01440625\\
3	0.00392857142857143\\
3.5	0.000618421052631579\\
4	7.89473684210526e-05\\
4.5	8.92857142857143e-06\\
5	0\\
5.5	0\\
6	0\\
};
\addlegendentry{BP (100\%)}

\addplot [color=blue, dashed, mark=+, mark options={solid, blue}]
  table[row sep=crcr]{%
2	0.2115\\
2.5	0.05775\\
3	0.00896451612903226\\
3.5	0.000858803986710963\\
4	6.09230769230769e-05\\
};
\addlegendentry{BP (3\%) $f\!=\!1/2$}

\addplot [color=blue, dashed, mark=o, mark options={solid, blue}]
  table[row sep=crcr]{%
2	0.117458333333333\\
2.5	0.02625\\
3	0.00382461240310078\\
3.5	0.00042158273381295\\
4	4.09472126295017e-05\\
4.5	3.01825154463461e-06\\
};
\addlegendentry{BP (3\%) $f\!=\!1/4$}

\addplot [color=blue, dashed, mark=diamond, mark options={solid, blue}]
  table[row sep=crcr]{%
2	0.11015625\\
2.5	0.026125\\
3	0.00500342465753425\\
3.5	0.000797717842323651\\
4	0.000105193075898802\\
4.5	2.77777777777778e-05\\
};
\addlegendentry{BP (3\%) $f\!=\!1/8$}

\addplot [color=magenta, dashed, mark=asterisk, mark options={solid,
magenta}]
  table[row sep=crcr]{%
2	0.471916666666667\\
2.5	0.194479166666667\\
3	0.0475166666666667\\
3.5	0.00773333333333333\\
4	0.0005\\
};
\addlegendentry{BP (3\%) random}

\end{axis}
\end{tikzpicture}%
}}
	$\quad$
	\subfloat[\label{fig:awgn25} Block-error rate versus $E_b/N_0$ in
	AWGN for
	$\mathrm{RM}(2,5)$ with $w=0.2$ and $f= 1/4$]{\scalebox{0.74}{
%
%

\pgfplotsset{
  every axis plot/.append style={very thick}
}

\begin{tikzpicture}

\begin{axis}[%
width=2.5in,
height=2.1in,
scale only axis,
xmin=2.5,
xmax=4,
xlabel style={font=\color{white!15!black}},
ymode=log,
ymin=0.001,
ymax=0.05,
yminorticks=true,
ylabel style={font=\color{white!15!black}},
axis background/.style={fill=white},
title style={font=\bfseries},
xmajorgrids,
ymajorgrids,
yminorgrids,
legend columns=2,
legend style={nodes={scale=0.8, transform shape}, anchor=south west,at={(0.03,0.03)},legend cell align=left, align=left,/tikz/column 2/.style={column sep=0pt}}
]

\addplot [color=red]
  table[row sep=crcr]{%
0	0.308453922315308\\
0.5	0.228297632468997\\
1	0.152513650913199\\
1.5	0.0977316602316602\\
2	0.0567544843049327\\
2.5	0.0288996717568146\\
3	0.0126576344287656\\
3.5	0.00526634167170765\\
4	0.00181481481481481\\
};
\addlegendentry{MRB $\nu=3$}

\addplot [orange,only marks,mark=diamond]
  table[row sep=crcr]{%
4	0.02\\
};
\addlegendentry{\cite[Fig.~3, Alg.~II]{Bossert-it86}}

\addplot [color=green!50!black]
  table[row sep=crcr]{%
0	0.326904761904762\\
0.5	0.234285714285714\\
1	0.157594936708861\\
1.5	0.102610837438424\\
2	0.0560200668896321\\
2.5	0.0291141732283465\\
3	0.0134928848641656\\
3.5	0.00557119205298013\\
4	0.00187082405345212\\
};
\addlegendentry{LP (100\%)}

\addplot [color=blue]
  table[row sep=crcr]{%
0	0.337440476190476\\
0.5	0.247380952380952\\
1	0.168291139240506\\
1.5	0.114137931034483\\
2	0.0662374581939799\\
2.5	0.0365354330708661\\
3	0.0174644243208279\\
3.5	0.00820364238410596\\
4	0.00293986636971047\\
};
\addlegendentry{BP (100\%)}

\addplot [color=green!50!black, only marks, mark=asterisk, mark
options={solid, green!50!black}]
  table[row sep=crcr]{%
0	0.335595238095238\\
0.5	0.242083333333333\\
1	0.162721518987342\\
1.5	0.106896551724138\\
2	0.058428093645485\\
2.5	0.0305511811023622\\
3	0.0140232858990944\\
3.5	0.0058112582781457\\
4	0.00195991091314031\\
};
\addlegendentry{LP (20\%)}

\addplot [color=blue, dashed, mark=asterisk, mark options={solid, blue}]
  table[row sep=crcr]{%
0	0.368988095238095\\
0.5	0.262916666666667\\
1	0.178291139240506\\
1.5	0.12\\
2	0.0673244147157191\\
2.5	0.0356102362204724\\
3	0.0167141009055627\\
3.5	0.00752483443708609\\
4	0.00251670378619154\\
};
\addlegendentry{BP (20\%)}

\addplot [color=green!50!black, dashed, mark=*, mark options={solid,
green!50!black}]
  table[row sep=crcr]{%
0	0.389880952380952\\
0.5	0.287916666666667\\
1	0.202025316455696\\
1.5	0.136206896551724\\
2	0.0774247491638796\\
2.5	0.0418700787401575\\
3	0.0202716688227684\\
3.5	0.00864238410596026\\
4	0.0028173719376392\\
};
\addlegendentry{LP (10\%)}

\addplot [color=blue, dashed, mark=*, mark options={solid, blue}]
  table[row sep=crcr]{%
0	0.526071428571429\\
0.5	0.385297619047619\\
1	0.266075949367089\\
1.5	0.172167487684729\\
2	0.0982441471571906\\
2.5	0.0509448818897638\\
3	0.0242820181112549\\
3.5	0.0105546357615894\\
4	0.00364142538975501\\
};
\addlegendentry{BP (10\%)}

\end{axis}
\end{tikzpicture}%
}}
	\caption{\label{fig:mixed3} Results of three numerical performance comparisons }
\end{figure*}

\subsection{Bit-Flipping Decoding}


BF is an iterative decoding method for binary linear codes transmitted
over the BSC~\cite{RU-2008}.  In its simplest form, it is based on
flipping a codeword bit that maximally reduces the number of PC
equations that are currently violated.  In one case, we also compare with the weighted BF (WBF) decoder proposed in~\cite{Bossert-it86}. This extends the idea to general channels by including weights and thresholds to decide which bits to flip.



\subsection{Most-Reliable Basis Decoding}

%


MRB decoding, which was introduced by Dorsch in
1974~\cite{Dorsch-it74}, is based on sorting the received vector
according to reliability (similar to Algorithm~\ref{alg:goodbad}).  After sorting, it
uses linear algebra to find an information set (i.e., a set of
positions whose values determine the entire codeword) containing the
most reliable bits. Then, it assumes there are at most $\nu$ errors
in the $k$ reliable positions. Then, one can encode the information
set implied by each of these error patterns and generate a list of
$\binom{k}{\nu}$ candidate codewords. Finally, the ML decoding rule is
used to choose the most-likely candidate codeword.

\section{Numerical Results}

For the numerical results, we consider various RM codes with length
$n = 32$ and $n = 128$. The code parameters are summarized in
Table~\ref{tab:rm_codes}. For all data points, at least 100 codeword
errors were recorded.

\begin{table}[b]
	\caption{Code parameters}
	\centering
	\renewcommand{\arraystretch}{1.3}
	\begin{tabular}{ c | c | c | c | c | c | c}
		code & $n$ & $k$ & $\dmin$ & $d^{\perp}_{\min}$ & rate & $F(r,m)$ \\\hline \hline
		RM$(2,5)$ & $32$ & $16$ & $8$ & $8$ & 0.5 & 620 \\
		RM$(2,7)$ & $128$ & $29$ & $32$ & $8$ & 0.23 & 188,976 \\
		RM$(3,7)$ & $128$ & $64$ & $16$ & $16$ & 0.5 & 94,488 \\
		RM$(4,7)$ & $128$ & $99$ & $8$ & $32$ & 0.77 & 10,668 \\\hline
	\end{tabular}
	\label{tab:rm_codes}
\end{table}

\subsection{The Binary Erasure Channel}

%
%
%

For a linear code on the BEC, the complexity of ML decoding is at
most cubic in the block-length~\cite{RU-2008}. Still, the BEC provides
a useful proving ground for general iterative-decoding schemes.  In
this section, we evaluate the PD for RM codes with
redundant PC matrices derived from the complete set of MWPCs.

%

For brevity, we focus only on the rate-$1/2$ codes $\mathrm{RM}(2,5)$
and $\mathrm{RM}(3,7)$. Fig.~\ref{fig:bec12} shows simulation results
for ML decoding, the PD with all MWPCs ($100$\%), and the PD when the
PC matrix is tailored to the received sequence using a fixed fraction
of available MWPCs. Note that for the BEC, the sets $G$ and $B$ used
in Algorithm 2 are simply the unerased and erased bits, i.e., the LLR
sorting is not employed. 

For the shorter code, all three curves are quite close together.
For the longer code, the PD matches the ML decoder when the
full set of MWPCs is used.
The performance loss of the low-complexity decoder is relatively small for the range tested.


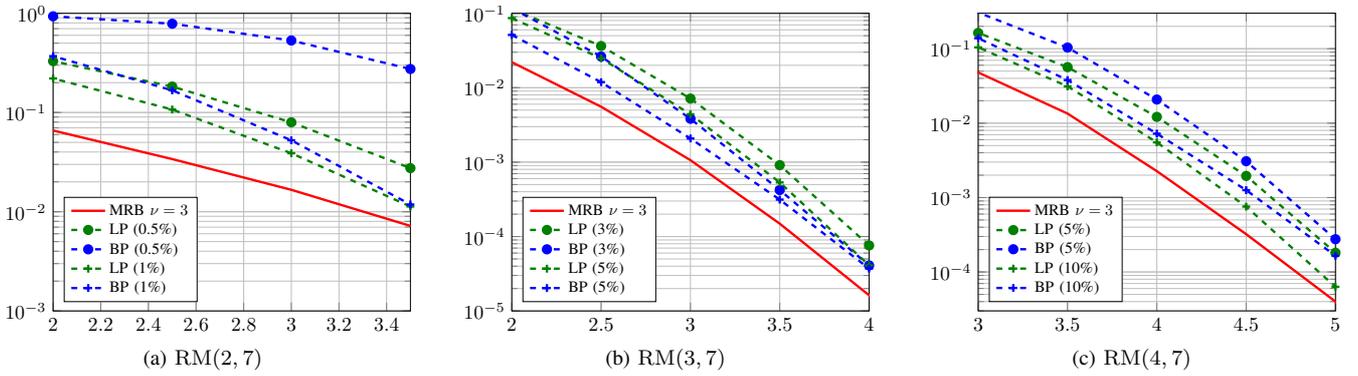
\begin{figure*}
	\centering
	\subfloat[$\mathrm{RM}(2,7)$]{\scalebox{0.74}{
%
%
\pgfplotsset{
  every axis plot/.append style={very thick}
}
\begin{tikzpicture}

\begin{axis}[%
width=2.5in,
height=2.1in,
scale only axis,
xmin=2,
xmax=3.5,
xlabel style={font=\small},
ymode=log,
ymin=0.001,
ymax=1,
yminorticks=true,
ylabel style={font=\small},
axis background/.style={fill=white},
title style={font=\bfseries},
xmajorgrids,
ymajorgrids,
yminorgrids,
legend style={nodes={scale=0.8, transform shape}, anchor=south west,at={(0.03,0.03)},legend cell align=left, align=left}
]
\addplot [color=red]
  table[row sep=crcr]{%
0.5	0.274207369323051\\
1	0.199625701809108\\
1.5	0.110079119367045\\
2	0.0658571722576662\\
2.5	0.0339496817217339\\
3	0.0166384407404106\\
3.5	0.0071958339908474\\
4	0.00254391222587744\\
4.5	0.000833252165980506\\
5	0.000229166666666667\\
5.5	5.66666666666667e-05\\
};
\addlegendentry{MRB $\nu=3$}

\addplot [color=green!50!black, dashed, mark=*, mark options={solid,
green!50!black}]
  table[row sep=crcr]{%
0.5	0.812339331619537\\
1	0.690580162195883\\
1.5	0.510835913312694\\
2	0.329285861288331\\
2.5	0.182661412549257\\
3	0.0796565350447158\\
3.5	0.0274998684971858\\
4	0.00649246663402894\\
4.5	0.00113139371075756\\
5	0.000115416666666667\\
5.5	2.16666666666667e-05\\
};
\addlegendentry{LP (0.5\%)}

\addplot [color=blue, dashed, mark=*, mark options={solid, blue}]
  table[row sep=crcr]{%
0.5	0.999143101970866\\
1	0.99812850904554\\
1.5	0.988648090815273\\
2	0.932702202099197\\
2.5	0.785450136404971\\
3	0.531895296669341\\
3.5	0.274662037767608\\
4	0.0978628392073609\\
4.5	0.0230174775087245\\
5	0.00339791666666667\\
5.5	0.000309583333333333\\
};
\addlegendentry{BP (0.5\%)}

\addplot [color=green!50!black, dashed, mark=+, mark options={solid,
green!50!black}]
  table[row sep=crcr]{%
0.5	0.721508140531277\\
1	0.559575795383656\\
1.5	0.385276917784658\\
2	0.220621527063182\\
2.5	0.107183995150045\\
3	0.0388923552307098\\
3.5	0.0112882015675135\\
4	0.00224193703359703\\
4.5	0.000305268434293642\\
5	1.95833333333333e-05\\
5.5	0\\
};
\addlegendentry{LP (1\%)}

\addplot [color=blue, dashed, mark=+, mark options={solid, blue}]
  table[row sep=crcr]{%
0.5	0.959725792630677\\
1	0.852776044915783\\
1.5	0.636050911592707\\
2	0.369829182959457\\
2.5	0.167323431342831\\
3	0.0525002228362599\\
3.5	0.0118247330492873\\
4	0.00150987596140208\\
4.5	0.000139568253033085\\
5	6.66666666666667e-06\\
5.5	0\\
};
\addlegendentry{BP (1\%)}

\end{axis}
\end{tikzpicture}%
}}
	$\quad$
	\subfloat[$\mathrm{RM}(3,7)$]{\scalebox{0.74}{
%
%
\pgfplotsset{
  every axis plot/.append style={very thick}
}

\begin{tikzpicture}

\begin{axis}[%
width=2.5in,
height=2.1in,
scale only axis,
xmin=2,
xmax=4,
xlabel style={font=\small},
ymode=log,
ymin=1e-05,
ymax=0.1,
yminorticks=true,
ylabel style={font=\small},
axis background/.style={fill=white},
title style={font=\bfseries},
xmajorgrids,
ymajorgrids,
yminorgrids,
legend style={nodes={scale=0.8, transform shape}, anchor=south west,at={(0.03,0.03)},legend cell align=left, align=left}
]
\addplot [color=red]
  table[row sep=crcr]{%
2	0.0220348039728752\\
2.5	0.00555361950209024\\
3	0.00106235273475712\\
3.5	0.00014895\\
4	1.625e-05\\
4.5	1.12128146453089e-06\\
};
\addlegendentry{MRB $\nu = 3$}

\addplot [color=green!50!black, dashed, mark=*, mark options={solid,
green!50!black}]
  table[row sep=crcr]{%
2	0.114208333333333\\
2.5	0.0363333333333333\\
3	0.00715213178294574\\
3.5	0.000911151079136691\\
4	7.58756783423779e-05\\
4.5	3.16028691144095e-06\\
};
\addlegendentry{LP (3\%)}

\addplot [color=blue, dashed, mark=*, mark options={solid, blue}]
  table[row sep=crcr]{%
2	0.117458333333333\\
2.5	0.02625\\
3	0.00382461240310078\\
3.5	0.00042158273381295\\
4	4.09472126295017e-05\\
4.5	3.01825154463461e-06\\
};
\addlegendentry{BP (3\%)}

\addplot [color=green!50!black, dashed, mark=+, mark options={solid,
green!50!black}]
  table[row sep=crcr]{%
2	0.0861666666666667\\
2.5	0.0248541666666667\\
3	0.00440116279069767\\
3.5	0.000533812949640288\\
4	4.10458806117415e-05\\
4.5	1.34933598466018e-06\\
};
\addlegendentry{LP (5\%)}

\addplot [color=blue, dashed, mark=+, mark options={solid, blue}]
  table[row sep=crcr]{%
2	0.0512291666666667\\
2.5	0.011875\\
3	0.00207945736434109\\
3.5	0.000313669064748201\\
4	3.7099161322151e-05\\
4.5	2.69867196932036e-06\\
};
\addlegendentry{BP (5\%)}

\end{axis}
\end{tikzpicture}%
}}
	$\quad$
	\subfloat[$\mathrm{RM}(4,7)$]{\scalebox{0.74}{
%
%

\pgfplotsset{
  every axis plot/.append style={very thick}
}

\begin{tikzpicture}

\begin{axis}[%
width=2.5in,
height=2.1in,
scale only axis,
xmin=3,
xmax=5,
xlabel style={font=\small},
ymode=log,
ymin=3e-05,
ymax=0.3,
yminorticks=true,
ylabel style={font=\small},
axis background/.style={fill=white},
title style={font=\bfseries},
xmajorgrids,
ymajorgrids,
yminorgrids,
legend style={nodes={scale=0.8, transform shape}, anchor=south west,at={(0.03,0.03)},legend cell align=left, align=left}
]
\addplot [color=red]
  table[row sep=crcr]{%
1	0.740825688073395\\
1.5	0.5136\\
2	0.279373368146214\\
2.5	0.129577464788732\\
3	0.0479688202668266\\
3.5	0.01345272627906\\
4	0.00226465301269621\\
4.5	0.000322226910176228\\
5	4e-05\\
5.5	1.66666666666667e-06\\
6	0\\
};
\addlegendentry{MRB $\nu = 3$}

\addplot [color=green!50!black, dashed, mark=*, mark options={solid,
green!50!black}]
  table[row sep=crcr]{%
1	0.903669724770642\\
1.5	0.7696\\
2	0.562228024369017\\
2.5	0.332394366197183\\
3	0.163243891470544\\
3.5	0.0565855299112961\\
4	0.0121300476992541\\
4.5	0.00195350064294338\\
5	0.00018125\\
5.5	1.33333333333333e-05\\
6	1.66666666666667e-06\\
};
\addlegendentry{LP (5\%)}

\addplot [color=blue, dashed, mark=*, mark options={solid, blue}]
  table[row sep=crcr]{%
1	0.997706422018349\\
1.5	0.9664\\
2	0.866840731070496\\
2.5	0.591549295774648\\
3	0.30700044970769\\
3.5	0.103838230966494\\
4	0.0207711143508231\\
4.5	0.00307928091037158\\
5	0.000275625\\
5.5	2.41666666666667e-05\\
6	8.33333333333333e-07\\
};
\addlegendentry{BP (5\%)}

\addplot [color=green!50!black, dashed, mark=+, mark options={solid,
green!50!black}]
  table[row sep=crcr]{%
1	0.853211009174312\\
1.5	0.696\\
2	0.468233246301131\\
2.5	0.23943661971831\\
3	0.104182281517014\\
3.5	0.0312355488291924\\
4	0.00552009171844701\\
4.5	0.000756226279819835\\
5	6.3125e-05\\
5.5	5.83333333333333e-06\\
6	4.16666666666667e-07\\
};
\addlegendentry{LP (10\%)}

\addplot [color=blue, dashed, mark=+, mark options={solid, blue}]
  table[row sep=crcr]{%
1	0.981651376146789\\
1.5	0.8768\\
2	0.646649260226284\\
2.5	0.344466800804829\\
3	0.138509968520462\\
3.5	0.0379198721991003\\
4	0.00722565851863385\\
4.5	0.00125869886787589\\
5	0.000165\\
5.5	1.54166666666667e-05\\
6	8.33333333333333e-07\\
};
\addlegendentry{BP (10\%)}

\end{axis}
\end{tikzpicture}%
}}
	\caption{\label{fig:awgn2347} Block-error rate versus $E_b/N_0$ in
	AWGN with
	$w = 0.05$ and $f=1/4$. }
	\vspace{-4mm}
\end{figure*}


\subsection{The Binary-Input Additive White Gaussian Noise Channel}
\label{sec:strategy}




For the binary-input AWGN channel, we consider both LP and BP
decoding. For the LP decoding, the ADMM solver is employed with
parameters $\mu = 0.03$ and $T_{\text{max}} = 1000$. For BP, we
perform $\ell = 30$ iterations and the weighting factor $w$ is
optimized for each scenario. As a comparison, we use MRB with $\nu =
3$ to approximate ML performance. 

First, we fix the number of rows in $\Hsub$ and illustrate how
different strategies for picking MWPCs affect the performance under BP
decoding. To that end, the code RM(3,7) is considered with $s = 2835$,
i.e., $\Hsub$ contains $s/F(3,7) \approx 3$\% of the complete set of
MWPCs.  Simulation results are presented in Fig.~\ref{fig:awgn_opt_f}.
The performance is shown for three different values for the parameter
$f$.  The proposed tailoring strategy leads to better performance
compared to picking a random set of MWPCs, with a performance gain up
to around 0.5 $\mbox{dB}$ at a block-error rate of $10^{-3}$. It can
also be seen that for an optimized choice of $f = 1/4$, the tailoring
strategy leads to a better performance compared to using the full set
of MWPCs.  This can be attributed to the reduced number of cycles in
the Tanner graph for $\Hsub$ compared to $\Hfull$. In the following,
we fix $f = 1/4$ for all other simulations, noting that it may be
possible to increase performance by re-optimizing $f$ for each
considered case. We also remark that for LP decoding, similar
observations regarding the optimal value of $f$ can be made and the
results are omitted.

Simulation results for RM$(2,5)$ using LP and BP decoding are shown in
Fig.~\ref{fig:awgn25}. For this code, LP decoding outperforms BP
decoding and gives virtually identical block-error rates as MRB
decoding using both $\Hfull$ and $\Hsub$ with $20$\% of available
MWPCs. For this case, it can be seen again that BP decoding benefits
from using a sub-sampled PC matrix $\Hsub$ compared to using $\Hfull$.
Also, a simulation point based on WBF is included
from~\cite{Bossert-it86} to show the superiority of LP and BP decoding.
Comparing with \cite[Fig.~5]{Hehn-isit07}, these curves nearly match
the ML results for the [31,16,7] BCH code and our ``BP
(10\%)'' result is quite close to their ``MBBP
$l=6$'' curve.

Lastly, we study the performance of three RM codes with length $n =
128$ and a range of rates.
The performance is shown in Fig.~\ref{fig:awgn2347} for a varying number of rows in $\Hsub$. From these graphs, we see that the
best performance is achieved at a different fraction of rows for each code. In particular, one requires around 1890 rows for RM$(2,
7)$ ($1\%$ of $\Hfull$), $4724$ rows for RM$(3,7)$ ($5\%$ of
$\Hfull$), and $1067$ rows for RM$(4,7)$ ($10\%$ of $\Hfull$).
These values were chosen so that the performance of the best scheme was roughly 0.25 dB from MRB at a block-error rate of $10^{-2}$.

\subsection{The Binary Symmetric Channel}
\label{sec:bsc}

\begin{figure}
\centering
\scalebox{0.74}{
%
%
\definecolor{mycolor1}{rgb}{1.00000,0.00000,1.00000}%

\pgfplotsset{
  every axis plot/.append style={very thick}
}

\begin{tikzpicture}

\begin{axis}[%
width=2.5in,
height=2.1in,
scale only axis,
scaled x ticks = false,
x tick label style={/pgf/number format/fixed},
xmin=0.02,
xmax=0.09,
ymode=log,
ymin=0.001,
ymax=0.1,
yminorticks=true,
axis background/.style={fill=white},
xmajorgrids,
ymajorgrids,
yminorgrids,
legend style={nodes={scale=0.8, transform shape}, at={(0.97,0.03)}, anchor=south east, legend cell align=left, align=left}
]
\addplot [color=red, solid]
  table[row sep=crcr]{%
0.02	0.00183333333333333\\
0.03	0.00705\\
0.04	0.0226\\
0.05	0.04375\\
0.06	0.0809\\
0.07	0.1174\\
};
\addlegendentry{RM(2,5) ML}

\addplot [color=green!50!black, only marks, mark=o, mark
options={solid, green!50!black}]
  table[row sep=crcr]{%
0.02	0.0026\\
0.03	0.00955\\
0.04	0.02305\\
0.05	0.0449\\
0.06	0.0771\\
0.07	0.1245\\
};
\addlegendentry{RM(2,5) BF}

\addplot [color=blue, dashed]
  table[row sep=crcr]{%
0.02	0.002\\
0.03	0.00875\\
0.04	0.02475\\
0.05	0.04845\\
0.06	0.0837\\
0.07	0.1268\\
};
\addlegendentry{RM(2,5) BP}

\addplot [color=orange, only marks, mark=diamond]
  table[row sep=crcr]{%
0.053671307382389	0.06\\
0.0354019062566282	0.017\\
0.0213215673911959	0.0032\\
};
\addlegendentry{BCH(31,16) BF}

\addplot [color=green!50!black, dashed, mark=+, mark options={solid,
}]
  table[row sep=crcr]{%
0.04	0.0018\\
0.05	0.0065\\
0.06	0.0275\\
0.07	0.0625\\
0.08	0.159\\
};
\addlegendentry{RM(3,7) BF}

\addplot [color=blue, dashed, mark=+, mark options={solid,
}]
  table[row sep=crcr]{%
0.04	0.003\\
0.05	0.01375\\
0.06	0.03725\\
0.07	0.0775\\
0.08	0.1725\\
};
\addlegendentry{RM(3,7) BP}

\end{axis}
\end{tikzpicture}%
}
\vspace{-2mm}
\caption{\label{fig:bsc12} Block-error rate versus channel-error rate on the BSC for $w=0.08$}
\end{figure}
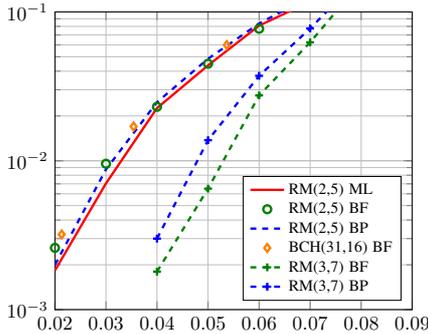

While the BEC reveals the locations where bits are lost and the
binary-input AWGN channel gives soft information for each bit, the BSC
provides no indication of reliability for received bits.  As we saw in
the last section, the performance in AWGN actually improves when a
subset of more reliable MWPCs are used for BP decoding. On the BSC,
however, it is not possible to advantageously select PCs for BP
decoding. Similarly, decoders based on ordered statistics provide no
gains on the BSC. Therefore, MRB cannot be used to provide a reference
curve for the approximate ML performance.

We consider both BP and BF decoding for the BSC using $\Hfull$, i.e.,
the full set of MWPCs. Fig.~\ref{fig:bsc12} shows our simulation
results for $\mathrm{RM}(2,5)$ and $\mathrm{RM}(3,7)$. The BF decoder
is identical to Algorithm II in~\cite{Bossert-it86}. For
$\mathrm{RM}(2,5)$, both the BP and BF decoder perform very close to
the ML decoder. The figure also includes results from
\cite{Bossert-it86} for the $[31,16,7]$ BCH code under BF decoding.
One can see that the results for the $[31,16,7]$ BCH and the
$[32,16,8]$ RM code are very similar. This is not surprising
because the two codes are nearly identical, i.e., the $[32,16,8]$ eBCH
code is equivalent to the code $\mathrm{RM}(2,5)$. Interestingly, for
the longer code $\mathrm{RM}(3,7)$, the BF decoder outperforms the BP
decoder with an optimized weight factor.

\section{Conclusions}

We have investigated the iterative decoding of RM codes based on
redundant PC matrices whose rows contain MWPCs.
Various iterative schemes were considered for the BEC, binary-input
AWGN channel, and the BSC. For the $[32,16,8]$ code RM$(2,5)$, near-ML
performance can be achieved using the PD on the BEC, LP decoding on
the AWGN channel, and BF or BP decoding on the BSC. For RM codes with
$n = 128$ on the BEC, the PD remained very close to optimal. For the
AWGN channel, the performance gap of LP and BP decoding with respect
to ML decoding increases. It was also shown that, for all channels
with the exception of the BSC, the complexity can be reduced by using
only a fraction of the available MWPCs. For BP, this strategy also
translates into better performance by reducing cycles. 


%

\end{document}